# DATA, TEXT, AND WEB MINING FOR BUSINESS INTELLIGENCE: A SURVEY

Abdul-Aziz Rashid Al-Azmi

Department of Computer Engineering, Kuwait University, Kuwait
Fortinbras222@hotmail.com

## ABSTRACT

*The Information and Communication Technologies revolution brought a digital world with huge amounts of data available. Enterprises use mining technologies to search vast amounts of data for vital insight and knowledge. Mining tools such as data mining, text mining, and web mining are used to find hidden knowledge in large databases or the Internet. Mining tools are automated software tools used to achieve business intelligence by finding hidden relations, and predicting future events from vast amounts of data. This uncovered knowledge helps in gaining completive advantages, better customers' relationships, and even fraud detection. In this survey, we'll describe how these techniques work, how they are implemented. Furthermore, we shall discuss how business intelligence is achieved using these mining tools. Then look into some case studies of success stories using mining tools. Finally, we shall demonstrate some of the main challenges to the mining technologies that limit their potential.*

## KEYWORDS

*business intelligence, competitive advantage, data mining, information systems, knowledge discovery*

## 1. INTRODUCTION

We live in a data driven world, the direct result of advents in information and communication technologies. Millions of resources for knowledge are made possible thanks to the Internet and Web 2.0 collaboration technologies. No longer do we live in isolation from vast amounts of data. The Information and Communication Technologies revolution provided us with convenience and ease of access to information, mobile communications and even possible contribution to this amount of information. Moreover, the need of information from these vast amounts of data is even more pressing for enterprises. Mining information from raw data is an extremely vital and tedious process in today's information driven world. Enterprises today rely on a set of automated tools for knowledge discovery to gain business insight and intelligence. Many branches of knowledge discovery tools were developed to help today's competitive business markets thrive in the age of information. World's electronic economy has also increased the pressure on enterprises to adapt to such new business environment. Main tools for getting information from these vast amounts are automated mining tools, specifically speaking data mining, text mining, and web mining.

Data Mining (DM) is defined as the process of analysing large databases, usually data warehouses or internet, to discover new information, hidden patterns and behaviours. It's an automated process of analysing huge amounts of data to discover hidden traits, patterns and to predict future trends and forecast possible opportunities. DM analyse datasets of rational databases, in multiple dimensions and angles, producing a summary of the general trends found in the dataset, relationships and models that fits the dataset. DM is a relatively new interdisciplinary field involving computer science, statistical modelling, artificial intelligence, information science, and machine learning [1]. One of the main uses of DM is business intelligence and risk management [2]. Enterprises must make business critical decisions based on large datasets stored in their databases, DM directly affect decision-making. DM is relied on in retail, telecommunication, investment, insurance, education, and healthcare industries they are data-driven. Other uses of DM includes biological research such as DNA and the human genome project, geospatial and weather research for analysing raw data used to analyse geological phenomenon.

A related field is Text Mining (TM), which deals with textual data rather than records. TM is defined as automatic discovery of hidden patterns, traits, or unknown information from textual data [7]. Textual data makes up huge amounts of data found on World Wide Web WWW, aside from multimedia. TM is related field to DM, but differs in its techniques and methodologies used. TM is also an interdisciplinary field encompassing computational linguistics, statistics, and machine learning. TM uses complex Natural Language Processing (NLP) techniques. It involves a training period for the TM tool to comprehend patterns and hidden relations. The process of mining text documents involve linguistically and semantically analysis of the plain text, thus structuring the text. Finally relates and induces some hidden traits found in the text, like frequency of use for some words, entity extractions, and documents summarizations. TM is used, aside from business applications, for scientific research, specifically medical and biological [22]. TM is very useful in finding and matching proteins' names and acronyms, and finding hidden relations between millions of documents.

The other mining technique is Web Mining (WM). WM is defined as automatic crawling and extraction of relevant information from the artefacts, activities, and hidden patterns found in WWW. WM is used for tracking customers' online behaviour, most importantly cookies tracking and hyperlinks correlations. Unlike search engines, which send agents to crawl the web searching for keywords, WM agents are far more intelligent. WM work by sending intelligent agents to certain targets, like competitors sites' [8]. These agents collect information from the host web server and collect as much information from analysing the web page itself. Mainly they look for the hyperlinks, cookies, and the traffic patterns. Using this collected knowledge enterprises can establish better customer relationships, offers and target potential buyers with exclusive deals. The WWW is very dynamic, and web crawling is repetitive process where contentious iteration will achieve effective results. WM is used for business, stochastic, and for criminal and juridical purposes mainly in network forensics.

In this survey paper, we shall look at the main mining technologies used through information systems for business applications to gain new levels of business intelligence. Furthermore, we shall look at how these techniques can help in achieving both business leadership and risk management by illustrating real enterprises' own experience using mining techniques. In addition, we shall look at the main challenges facing data, web, and text mining today.

## 2. HISTORY AND BACKGROUND

Many developments were made leading to mining technologies we have today. These developments date back to early days of mathematical models and statically analysis using regression and Bayesian methods in mid-1700s. With the advent of commercial electronic computers after World War II, large data sets were stored into magnetic tapes to automate the work. In the 1960s were data stored in computers helped analysers to answer simple predictive questions. With the development of programming languages, specifically COmmon Business Oriented Language or COBOL, and Rational Database Management Systems RDBMS, querying databases were possible. Meaning more complex information and knowledge can be extracted. Development of advanced object oriented languages such as C++, Java, multi-dimensional databases, data warehousing, and Online Analytical Processing OLAP make way for an automated algorithmic way of extracting patterns, knowledge from such large data sets. DM tools today are more advanced and provide more than reporting capabilities, they can discover hidden patterns and knowledge. These DM tools were developed in the 1990s.

After the Internet and the WWW revolution in the early 1990s, many research and developments were made to automate the search and exploration of the net, especially text, found in the URLs. Developments in NLP, neural networks and text processing led ultimately to search engines development. The need for better search algorithms led to textual exploration of web pages. These developments greatly enhanced the search engines and opened the door for text mining to be applied in several other applications. Search engines' technologies were centred on agents that could map the vast WWW and correlate keywords and similar other possible keywords. These developments will lead to the more intelligent agents that search the WWW for not only keywords but also site visitors' patterns. Ultimately, the developments in both DM and TM lead to the notion of WM, were the WWW is used as a source for looking for new knowledge, hidden away somewhere. WM agents are small standalone software, that crawl the WWW, acquiring logging data, cookies, and site visits behaviour found on the servers and other machines attached to the WWW.

The tremendous advancements made in the mining technologies have shifted thought from data collection to knowledge discovery and collection [9]. With today's powerful and relatively inexpensive hardware and network infrastructure, matched with advanced software for mining, enterprises are adapting mining technologies as essential business processes. In addition, the Internet has an integral role as network and communications are ubiquitous today, mining is carried over the world through the network of databases. The vast amount of knowledge is not only consumed at the top senior management level but at all the other levels of an enterprise as well.

Today mining software utilizes complex algorithms for searching, pattern recognition, and forecasting complex stock market changes. IBM and Microsoft are on an epic race to produce best DM software to date; this is also influenced by security and intelligence agencies such as FBI and CIA. Multi-linguistic and semantic TM is a hot new research topic. As modern as it is today, WM has become an increasingly adopted business process as well. WM is suited more for e-commerce than DM and TM. The nature of e-commerce suggests the direct exploitation of customers' online behaviours. Many surveyors, such as Gartner Group, predict that over 5 billion dollars of business will be net worth of e-commerce in the coming years [10]. WM is heavily used for e-education and e-business, as the WWW is again their main platform. As developments were huge in the 1990's in terms of hardware support for mining techniques and the further leaps achieved by modern software, mining techniques are more of a must than a commonplace for modern business today. Relatively new and emerging mining techniques are what are known collectively as Reality Mining [65]. Reality mining is the collection of transactions made daily by individuals to realize how they live and react. Reality mining is aimed at developing our understanding of our modern societies, economies and politics. This is technology is made possible by the ICT world we live in today. Reality mining which is very controversial as it infiltrate individuals privacy, is catching the intention of governments and corporate, as it can be used for potential business benefits. Reality mining really mines what is known as reality traces, these include all patterns of human life in digital form. Traces include banking transactions, travel tickets, mobile telecommunications calls, blogs, and every possible digital transaction. The aim of such emerging technology is to better understand societies as well as individual and to further develop solutions aimed at them. The main problem facing such new mining technology is privacy concerns from individual, and governments, as data spread on the Internet is not really owned by any legislative body.

## 3. RELATED WORK

Much work was done in surveying business applications of the aforementioned mining techniques. However, most work considers each mining technique separate from one another. In [4] the authors have provided an overview of Knowledge Discovery in Databases (KDD) approaches. They also classified the approaches depending on software characteristics. In [5] the authors demonstrated how modern technologies shifted the process of decision-making, from manual data analysis using modelling and stochastic to an automated computer driven process. The authors also stated that knowledge discovery tools have benefits such as increased profitability. In addition, risk management and market segmentation is another advantage. A survey of visual data mining techniques is found in [11]. The authors have stated that large data sets with complex dimensions need a better way for representation. In their paper, the authors have reviewed previous work done in data visualization [12]. The authors classified data visualization techniques into six different classes, based on the parameters of the data.

In [13], the authors have surveyed the relatively young and interdisciplinary field of TM. Since most information found in computerized form are textual, the need to extracts this unstructured text into informative knowledge demands new tools. TM tools are machine tools that analyse written text with a certain context [15]. A case study of TM is found in [16], the paper discusses the use of TM for patent analysis. The authored discussed how professional patent information business is sceptical in using TM tools. The paper discussed showcased PackMOLE (Mining Online Expert on Packaging Patents), a TM tool, designed for mining patent information in the packaging field. The authors showed that PackMOLE tool has advantages over the manual patent portfolio analysis techniques. However, the tool calibration of its internal clustering processes is difficult, and consumes time. This leads to the use of a hybrid use of text mining techniques and manual patent classifications in conjunction. In [18], the authors presented a review of TM techniques. The authors clearly stated that TM faces challenges as natural language processing NLP techniques are not readily made for mining activities. The paper illustrated several TM technique that included information extraction, topic tracking, summarization, categorization, clustering, concept linkage, information visualization and question answering to name a few. Finally, the author stated that TM is used in media, banking, politics, and even in insurance.
How business intelligence is derived from web mining is found in [17]. WM or web usage mining is described as an intelligence tool to aid enterprises in the intense competition found in e-commerce. The paper presented a review of current WM techniques used as well as introducing a novel approach called intelligent miner. Intelligent miner (called i-Miner) is a hybrid framework for WM; it uses a combination of algorithms for finding and processing log files from web servers. Its then applies rules and structures to find hidden patterns found in the log files. In [20], the authors stated how difficulties arise in WB from all the fuzziness and unstructured nature of the data found in the WWW. The paper also illustrates the evolution of DM that lead to WM. The authors stated that WM has these main task, associations, classification, and sequential analysis. The paper included a WM study on two online courses. Using WM to improve the two online courses experience, based on the results from the WM tool that used the logging files. An excellent discussion on the characteristics of WM is found in [26]. The authors relate the development of the soft computing, which is a set of methodologies to achieve flexible and natural information processing capabilities. The paper discusses how it is difficult to mine the WWW with its unstructured, time varying, and fuzzy data. The paper also specifies four phases that include information retrieval IR, information extraction, generalization, and finally analysis of the gathered data. The authors also classified WM into three main categories, Web Content Mining WCM, Web Structured Mining WSM, and Web Usage Mining WUM. WCM is about retrieving and mining content found in the WWW like multimedia, metadata, hyperlinks, and text. WSM the mining of the structure of WWW, it finds all the relations regarding the hyperlinks structure, thus we can construct a map of how certain sites are formed, and the reason why some documents have more links than others. Finally, WUM, which is the mining of log files of web servers, browser generated logs, cookies, bookmarks and scrolls. WUM helps to find the surfing habits customers and provides insights on traffic of certain sites.

# 4. MINING FOR INFORMATION AND KNOWLEDGE

How does mining really work? Let's look on how DM works. It's regarded as the analysis step of the Knowledge Discovery in Databases KDD process [5]. KDD usually has three steps, pre-processing, then DM, then finally data verification [6]. For DM, it uses data stored in data warehouses for analysis. DM technologies use Artificial Intelligence AI and neural networks; a good review of neural networks applications in business is found in [45]. AI and neural networks are nonlinear predictive models [3] that learn through experience and training. Furthermore, AI techniques are highly used in business models and predictions [50]. AI led to the more advanced technique of machine learning. Machine learning is the ability of the machine to adapt and learns from previous trials and errors, and then it tries to find out how to be more effective. Other techniques are decision trees and genetic algorithms. Decision trees [21] are top down induction rationale thinking tools, they support classifying the decisions into different branches. Starting from their roots and stemming to the leaves, each decision branching out has risks, possibilities, and outcomes. Certain decision trees type used are Classification and Regression Trees CART.

DM tasks are several and depend on the different fields where the DM is applied. Classifying data stored in multi-dimensional databases is a prominent task. Classifying involves identifying all groups that can be found in the data, like grouping fraudulent transactions in a separate group from legitimate transactions. Associations and rule inductions, intelligently inferring if-then relations from patterns found hidden in the data. This leads to finding hidden correlations, like market baskets, which are the products bought mostly together. Another task is regression modelling or predictive modelling [47], which helps in predicting future trends. Usually regression is used for extrapolating data in mathematics, in DM; it helps to find a model that fits dataset. Data visualization, visually aids in linking multi-dimensional data together, such as Exploratory Data Analysis EDA and model visualization. Data visualization tasks are emerging task of newer interactive DM tools. Other tasks include Anomaly detection, were anomalies are caught. Summarizations that express information extracted in a compact form; and aggregation of data, where sums of data are compacted to single figures or graphs [4].

DM requires huge computational resources. DM requires as a perquisite a data source, usually a data warehouse, or a database. Data warehouses [19] are large databases used for data analysis and aggregation. It extracts and transforms the data from the DBMS the enterprise use for daily activates and through Extraction, Transformation, and Loading ETL processes [20]. ETL extracts the data from the DB, then it pre-process the data and finally load it into the data warehouse for further processing of this cleaned up data. Aside from this, DM requires substantial processing power, usually top class server level computing prowess. A growing interest in combing data mining with Cloud Computing technologies is emerging such as providing DM as a service, such as found in [40]. DM applies number crunching algorithms, parallel processing, and neural networks with AI techniques, thus requiring this huge computational resources. For standardizing DM, the CRoss Industry Standard Process for Data Mining or CRISP-DM for short, was developed [14]. CRISP-DM is standard model developed in 1997 by the ESPRIT funding initiative, as part of a Euro Union project, and lead by leading industry companies such as SPSS, now part of IBM, and NCR Corporation. Until data, CRISP-DM is the leading standard for DM, as most leading DM software implements it.

The next question is how does TM work? TM essentially is based on how do we read and comprehend text. This process of reading, then understanding what is read is to some extent imitated. However, this is not a very easy process for computers. For TM techniques, it first has to retrieve relevant documents, data, or text found on the WWW. For this step Information Retrieval IR systems are used [23], Google search engine is an example of such systems. The second process is NLP; this is the most difficult part of TM. In NLP, AI and neural networks are again used to parse the text the same way humans do to comprehend the text. The text is parsed; nouns and verbs are used to understand grammatically the meaning on each sentence. The final step is information extraction; here linguistic tools are used to get information from the comprehended text. Entities, characters, verbs, and places are correlated and hidden unknown new information is generated. This is where DM techniques are applied at the final stages to extract the information. This is how most TM tools work.

TM techniques try to emulate human comprehension of textual data. TM highly used in many applications to replace manual search in textual documents by humans. TM techniques, unlike DM techniques, deal much with unstructured data sets, thus more complicated. Healthcare and medical usage includes linking several hundred of medical records together, finding relations between symptoms and prescriptions [47]. Media applications also use TM, especially in the political aspects of certain controversial issues or controversial characters. TM is also used for clustering archived documents into several clusters according to predefined semantic categories [24]. TM is newly finding new uses in the legal and jurisdictional fields, as TM is applied to patents, and criminal profiling. TM is used in text summarizations, where it effectively identifies the main names, characters, verbs, and the most used words or referenced subjects in large documents. TM is also used in TM OLAP, found in [59], as a textual search tool rather than numerical. Furthermore, TM is an essential part of modern IR engines, such as Google's search engine and Yahoo's search engine, as they apply sophisticated TM techniques to correlate search queries together.

The final question is how does WM work? WM is based on Internet and agent technologies [25], that utilize soft computing and fuzzy logic techniques. Again, WM technologies rely on IR tools to find the data it needs. IR systems provide means and ways in which these intelligent agents can scour the WWW. In addition, it is worth noting that such IR systems are greatly supported by the developments of Semantic Web technologies, SW technologies [51]. SW technologies were developed to build semantics over web published content and information on the WWW, in the aim of easing the retrieval process of the content by humans and machines. As SW technologies are still a work in progress, as more new web technologies are implementing SW techniques to aid in the search process of such contents. The next step is the agents programming, these agents are programmed after finding their targets to apply their mining techniques. These include analysing the HTML document, parsing, and extracting all the hyperlinks, multimedia among other things. Users' preferences and online accounts are specifically tracked to identify their sessions and transactions logged in that system. Server data, like site traffic, activates and even possible proxies are retrieved for further analysis. Final step for these agents is to analyse the gathered data, using DM techniques, to understand the habits, patterns found in the WWW.

WM tasks depend on the mining purpose intended. WM is divided into three main categories as in [26]; web content mining WCM, web structure mining WSM, and web usage mining WUM. WCM is intended for retrieval can fetch and locate context sensitive text, multimedia, and hyperlinks depending on the giving context. Other WM tasks intended for WSM include finding the chain of links or site maps of certain sites, mainly to find were most of the traffic is headed. Finally, WM intended for WUM can collect logs, cookies, bookmarks, and even browsers history and metadata of targeted users. WUM is also used for mining social networks, namely online blogs [33]. These tasks are all WWW oriented, but WM also plays a key role in mining social media networks for investigations, intranets, and to lesser extent Virtual Private Networks VPNs. A new use of WM is multimedia mining, where WM tools will try to mine out multimedia as pictures, movies, audio files, and applications.

## 4.1. Mining Tools Software

Mining tools of sometimes called siftware, as they sift through the data. Mining tools varies depending on their sophistication, as state of the art tools are expensive. In 2008, the world market for business intelligence software reached over 7.8 billion USD according to [27]. IBM SPSS is an example of business intelligence software, offering mining capabilities. Clementine was a graphical and widely used DM tool of the late 1990's; this case study utilizes that software [43], it is precursor to SPSS. IBM also provides online services for WM, called Surfaid Analytics; they provide sophisticated tools for WM [32]. IBM is one of main providers of solution-oriented packages such as IBM's Cognos 8 solutions [44]. Other types of mining or business intelligence tools are integrated tools, like Oracle Data Mining, a part of the company flagship RDBMS software. SAS offers its SAS Enterprise Miner [47], a part of its enterprise solutions. Another major player in the enterprise and business information systems is SAP, it offers world known ERP solutions along with providing other mining tools software that can be integrated into their ERP solutions. Atos is an international information technology services provider that utilizes SAP software [63]. Microsoft offers SQL Server Analysis Services, a platform dependant solution integrated in Microsoft SQL for Microsoft Windows Server. Other Microsoft products include PowerPivot, a mining tool for small and middle size enterprises. Open source mining tools include the Waikato Environment for Knowledge Analysis or WEKA [28]. Other open source tools include RapidMiner and KNIME. Situation with open source mining tools however is not as other open source software, as their use is quite limited [49]. Furthermore, with huge decrease of in costs of storing and acquiring data from WWW, data acquisition tools such as RFID tag readers and imaging devices, e-commerce, and telecommunications, mining software costs of procuring dropped considerably [42].

A new trend in utilizing such mining tools for middle-sized and small enterprises is Cloud Computing based Mining tools [58]. As small and middle-sized enterprises, lack the infrastructure and budget available for large enterprises. Cloud Computing helps in providing such mining tools at relatively lower costs. Cloud Computing provides web data warehousing, were the actual data warehouse is outsourced and accessed through the web. They also provide sophisticated mining analysis of the data as the enterprise specifies and demands. Aside from lowering the costs of the mining tools infrastructure, Cloud based mining also provides expertise that are not available in such middle-sized and small enterprises. Usually start-up entrepreneur level enterprises lack not only the financial resources but also the human resources and expertise in the IT field. The Infrastructure-As-Service IAS provides middle-sized enterprises comfort from the burden of software, hardware, and human resources managements as well. The main downfalls of Cloud Computing are the dependency and privacy issues that occur from the fact that another party that the enterprise agrees to store its data on its machines and data warehouses. Such issues are limit and turn off large capable enterprises from going with Cloud based solution. These enterprises can set up their own mining solutions instead is much less risky. Dependency means that the whole service depends on the other party, not the enterprise itself, meaning that the enterprise is pretty much tied up with what the service provider has to offer. The privacy concerns arise from the fact that the enterprise's data is technically not under its control or even possession, the other party has it, it utilize its resources to give results and analysis. The privacy concern entails the misuse of the data, mostly causing confidentiality risks
.

## 5. BUSINESS INTELLIGENCE THROUGH DATA, TEXT, AND WEB MINING

Business applications rely on mining software solutions. Mining tools are now an integral part of enterprise decision-making and risk management. Acquiring information through mining is referred to as Business Intelligence BI. Enterprise datasets are growing rapidly, thanks to use of Information Systems IS, and data warehousing. On average, credit card companies usually have millions of transactions logged per year [29]. Largest data sets are usually generated by large telecommunications and mobile operators as they mount up to 100 million user accounts generating thousands of millions of data per year [30]. As these number mount up, analytical processing such as OLAP and manual comprehension seems ineffective. With BI such tasks are within reach. According to Gartner Group "Data mining and artificial intelligence are at the top five key technology areas that will clearly have a major impact across a wide range of industries within the next three to five years," this was back in a 1997 report [31]. Also according to Gartner Group reports in 2008 [53], it was found that 80% of data found in enterprises information systems is unstructured, and would very likely to double in size nearly every three months. BI has become the prevalent decision support systems in organizations. BI has dominated many industries including retail, banking, and insurance [41]. The First American Corporation FAC is an example of a success story in implementing BI to improve its customers' loyalty and better investment. It's worth mentioning that BI software is aimed at knowledge workers, mainly executives, analysts, middle management, and to a lesser extent operational management. Fig.1, illustrates how data today is transformed to acquire BI.

BI is implemented through mining tools; these tools generate findings that are ultimately used to gain competitive advantage over rivals, better and efficient business operations, and better survivability and risk management. Mining tools provide better customers' relationship management CMR, through mining real habits, patterns, and even customers churn. Customers churn is defined as the per cent of customers that have left the enterprise, most likely to other rivals or due to the inability to keep your competitive advantage and customer satisfaction levels. Habits and trends in customers' data help in discovering the customers' segmentations, what customers to target, especially alpha customers. Alpha customers are those that play a key role in a product success, thus finding what they want is essential. This means that, mining tools are essential for catalogue marketing industries and advertising agencies. In addition, mining tools, especially DM, provides market basket analysis that helps the discovery of products that are bought usually together. As modern economies around the world today are driven by information, becoming information and knowledge based economies [66]; BI tools are from the top reasons of development information technologies in business today. BI tools in business today are integrated in most enterprises tools such as Enterprise Resource Planning ERP tools, Customer Relationship Management CRM tools, supply chain management tools, data warehouses, and even RDBMS. BI is also the main tools for decision support in modern enterprises. BI tools provide competitive advantages, better customer relationships managements, and better management of risk in investments.

Mining tools provides predictive profiling; this means that using current and historical behaviours of your customers, possible future behaviours of purchase are predicted. The insurance industry is most interested in their customers' medical records and its history. Stock market predictions are mostly done using mining, NASDAQ, is a major DM user. NASDAQ had spent over 450 million dollars [64], estimated costs, on implementing a full BI solution for its stock market exchange. Their solution consists of smart BI agents that buy and sell stock autonomously, with human monitoring for erroneous errors. Google, the technology giant, uses BI on its Google Finance service [59]. The Google Finance web page contains dynamic charts of international stock markets, with references to critical points in the graph directing to web pages that are the service got its information, providing assurance to end users. Mining tools are also used for automatic spam detection, and the defence against fraud, through fraud detection techniques utilizing mining tools [30]. Most major banking and telecommunication companies apply automated fraud detection systems through mining techniques, AT&T, bank of America are examples of such users of fraud detection. In the next subsections we will look at the main aspects were BI through mining tools is used to gain business proficiency. BI and mining tools are used exchangeable in the following text.

## 5.1. Achieving Competitive Advantages

Competitive advantage is driven by competitive pressure. According to [41], competitive pressure is degree of pressure that companies feel from rivals and possible new entrants. This pressure is lessened by gaining a competitive advantage. For gaining competitive advantages, enterprises develop market research groups that analyse the large data sets to acquire knowledge. Market research, through mining, try and find what products dominate the market, why this is and what hidden elements that set such products leading in sales. For example, media networks use mining in their market research to set the common factors between audience and the program's scheduled slot. BBC, used to hire human experts to schedule its programs slots, now its uses fully automated mining tools for scheduling, the results were equivalent or better than the human manual scheduling [36]. Marketing use-mining tools to get the market's baskets, as mentioned before, market basket are associations of certain products that are highly likely bought together. No competitive retail enterprise is without its set of market baskets, leading in this segment are Wal-Mart, Costco, and K-Mart.

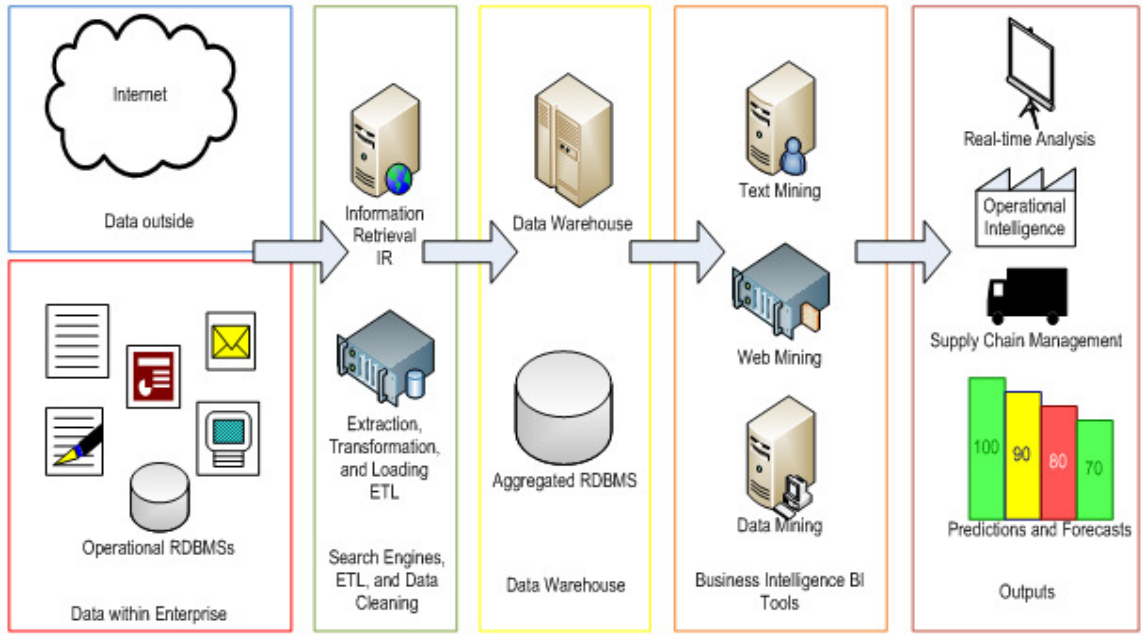

Figure 1. Data Transforming into Business Intelligence

For better risk management, in banking for example mining tools are used to automate risk of bankruptcy. An example is bankruptcy predictions [38]. Bankruptcy predictions are essential risk assessment processes, nearly all large banks such automated tools in their bankruptcy predictions. Most of these tools use neural networks in assessing the counterpart's liability to pay back the loan given. Bankruptcy prediction tools help the bank in reducing the risk of lending money to trouble or would be troubled customers. These tools also depending on the risk the bank is willing to take, will estimate the fair interest rates depending on the credibility of the other party. In addition, bankruptcy prediction tools are also used to assess the credit loss of current and on-going bank's loan portfolio. Credit loss is the prediction of possibility that depreciation or loss levels of loan will be more than the expected 1% in a predefined time frame. These tools came to the foreground after strict regulations especially after the 2008 market crash and following economic crisis that led many banks to file for bankruptcy. Banking sectors today utilize BI with bankruptcy tools together to create sophisticated profiles on their potential customers before any lending. Usually banks have setup return/risk parameters depending on the other party's market worth, market share, and current debt. According to [38], bankruptcy tools are also used by accounting companies. Auditing now takes into account the possibility of going bankrupt in the near future. Lawsuits are filed against accounting companies if they did not warn the contracting party of its possible bankruptcy. J.P. Morgan utilize its' own bankruptcy tool called CreditMetrics [39], it model the change in the credit quality ratings, it can obtain a possible estimate of the risk incurred in the loan.

In manufacturing industries, sophisticated Management Information Systems MIS are used to automate the processes. Such MIS tools generate reports that can be feed to BI tools to better optimize the manufacturing processes [61]. MIS tools are used to automate the work, but such tools were not meant to monitor the performance nor smartly detect how to use the operation data to further develop and optimize the manufacturing process. MIS software for the manufacturing industry includes Material requirements planning MRP systems, which detects when the finished products will be needed or shipped out of the factory. Other example of MIS software for manufacturing is Just in time JIT inventory systems, precise and immediate delivery of materials before usage. These tools also generate huge amount of raw data, reports, and schedules, which could be used for mining. According to [61], BI tools using such data can generate adaptive manufacturing schedules to match the demands. BI tools help the top management further to lessen the wastage and unnecessary expenditures. BI tools not only improve the manufacturing processes in the physical factory, but also optimize pricing analysis, by examining sales, production, and quarterly reports, the manufacturing enterprise can adjust its pricings on its items. Other optimization of other processes include warranty analysis, by examining the warranty claims and lawsuits occurring from manufacturing errors, failing parts, and possible risk factors. As such, the manufacturing enterprise can reduce such errors, forecast possible risks and avert flawed warranty strategies.

## 5.2. Improved Customers' Relationships

Customers today are well informed, and demanding. Meaning customers are getting even more difficult to please. For these reasons, Customer Relationship Management (CRM) systems were developed to better handle the customers. CRM is defined as the process of managing the relationships of a company's customers through technology focusing on the customers' satisfaction and further development of these relationships [46]. With information systems availability since the mid 1980's, companies have used information technologies further to develop their customers' relationships. CRM software automates how the company deals with its customers' information, but with the introduction of BI tools, further and newer exploitations of such information are possible. Questions like, how to keep your customers satisfied? What makes a customer happy? How can we attract potential customers? Why did some customers have our company left for other rival companies? All these questions are answered by BI tools. BI tools are used with existing CRM systems to help in customers oriented decisions. These decisions include customers' service, market targeting, product evaluation, and even in manufacturing. The main users of BI tools with CRM systems are retailers, electronic retailers or e-tailers, services providers, and telecommunications.

Though BI helps in gaining a competitive advantage, some businesses are heavily dependent on its customers' satisfaction [48]. This leads to extensive BI exploitation of the CRM data. Companies can achieve competitive advantages in providing value propositions to its customers; these value propositions include lowest cost, bundles and offers, special discounts, and loyalty and partnership privileges. Lowest cost however is not easily achieved, as the WWW had brought new ways for the enterprise customers to search for other rivals possibly offering better prices. Transparency in pricing is another factor that customers would appreciate and look for. BI tools help in avoiding price transparency problems by offering the enterprise the choice of price discrimination. Price discrimination [48] means that the company offers the same product with different prices for different people and regions; this of course depends on each individual's willingness to pay for how much. This dynamic trait allows the enterprise to appeal to its diverse customers with offers that are identical to the lowest cost offers. This ability is not possible without prior knowledge of how much each individual is willing to pay of course. Here is where BI tools help in evaluating your customers buying and spending habits. For example, some customers are willing to pay more for certain products, depending on personal preferences such as hobbies, age, or culture. These customers the company offers the price it thinks they will gladly pay, most of the time they will pay. Other customers, would never pay for that price, instead they may be willing to pay for less. Such customers the company can offer limited and special discount prices these customers will surely be happy with such pricings. It worth noting that sophisticated BI techniques must be employed to achieve effective price discrimination. A tool that utilizes DM in CRM software is found in [67], through a data mining engine DME, which mines the CRM database.

BI helps in assessing possible switching costs of customers. As stated in [48], high switching costs leads to lower customers' satisfaction. As customers are not satisfied with being stuck with a company because of such costs, it reduces their overall satisfaction. In addition, BI finds the market baskets, as stated earlier; this is from the customers' side though. As market basket, analysis can be considered from the customer side and the company market side. Through CRM data, BI tool can construct specific market baskets for certain groups of users that is more specific than the ordinary market baskets. Amazon, Google are among the best utilizes of customers market baskets. Market baskets are seen in the form or recommendations, special discounts on certain products, and the usually advertisements regarding specific products. Furthermore, market baskets implemented for e-tailers, another name for e-commerce retailers, are more effective because it is more easily exploited than traditional brick and mortar retailers, where certain products could not physically be with others.

Services industries have another use of BI tools; it is in their Customers Experience Management CEM systems. Such systems keep track of the experience of customers. BI tools help find evidence of what make customers satisfied, happy, and what do they expect from the enterprise. As rivals are always trying to take the enterprise's customers and recruiting them. BI is also used to find reasons for customers churns. Customers churn, as defined earlier, are enterprise health indicators. As high customers churn ratings are an alarming indicator that customers are unsatisfied. Churn analysis can also show reasons and causes for these turn over, like turning to rival companies, or found substitutes to the enterprise products. BI tools utilizing CEM systems can also help identify the alpha customers, as stated earlier, those highly valued customers that the company must take special care in dealing with them. The emerging Reality Mining RM techniques can also be used to acquire further insight of the costumers. As it is an emerging field, most of the methodologies are experimental. As mentioned before RM faces tremendous pressure as it deliberately violates individuals' privacy, anonymity, and even to a lesser extend security.

## 5.3. Better Logistics and Inventory Management

Supply chain management systems such as Warehouse Management Systems (WMS) [34] and Inventory Management Systems IMS; provide automated control and management of the supply chains. These systems achieve logistics and production efficiencies, such as high distributors and customers satisfactions, on demand supplies, and quick response to inventory levels. Logistics industries make profitability out of efficiency of their supply chain managements. Using BI tools, help supports supply chain management; enterprises have now better management over its supplies. BI tools, integrated in WMS or IMS, help find patterns, shortages, possible overproduction, and underproduction, but mostly quick response to demand spikes that are extremely important to catch.

Traditionally, complex mathematical models were used for supply chain managements and logistical problems. Logistics can affects inventories (under-flows or over-flows), specifically from slow response to spikes in demands and lack of precise forecasting. Order batching methods were used to batch many items to as many as possible different locations [35], also routing batched orders was another problem. Using association rules and clustering, BI can help in effectively in routing, and batch these orders. BI tools mine markets current situations, providing forecasting predictions. These predictions help in managing inventories, keeping under-flows and over-flows at bay. BI tools have complete control over forecasting, as demands and surges are estimated through reading market history, political events, or even rivals retributions. Modelling, beside prediction, is another BI facility used for warehouse management, as certain industries fall in certain demand and supply cycles. Models also help automate replenishment processes. It is worth noting that sophisticated BI tools can predict certain market segments demands. Such market segments are used in niche marketing.

BI help logistics in providing strict and timely decision supports. The utilization of using such BI tools results in delivering on time decisions and faster feedbacks, such benefits are of BI in food supply chain networks are found in [60]. In the food industry, time of delivering the goods is critical as the stock of food supplies or livestock can rot or die during anyone of the many supply chain stages. As poor management and slow reactions in such supply chains can give low yields, meaning that slow reactions can lead to massive loss that could be avoided. The problem of Death-On-Arrival DOA occurs when many of the livestock die during transport due to reasons unknown or known much later in the supply chain. BI tools can effective identify such signs of dying livestock at an early stage, that some of the livestock are dying; this information can help in reducing the casualties or identifying the problem and rescuing the remaining livestock. Aside from the livestock management and preservation in the supply chain, food products like milk, sugar, or fruits, and other raw food materials, spoil if not handled in time. BI help in dynamically routing the supplies before it rot during the storage, or distribution. Many BI tools provide timing managements of such supply chains.

One of the main users of BI tools in supply chain managements are the Third Party Logistics (TLP) or 3PL. TLPs are an outsourced party that provides logistics services to the contracting first party. TLPs using BI can provide much more services to their customers [62], in addition to the traditional services; BI enables new services such as detailed reports and forecasts about their delivered goods. Another additional service is the cost-benefit analysis, this service helps their customers to analyse their current supplies and evaluate their suppliers. Another service powered by BI tools is the supply chain visibility, allowing the customers to track their supplies online dynamically and in real time.

## 5.4. Anomalies and Fraud Detection

BI tools through its sifting capabilities can locate certain hidden patterns found in daily transactions. These patterns are used for possible fraud and anomaly detection. Fraud and anomaly detection is defined as detection of deceptive transactions or strange and unusual transactions that need inspection. BI tools are one of the main tools used in forensics to detect fraudulence. Most insurance and telecommunication companies use fraud detection daily. As it's the norm to investigate suspicious transaction for possible fraudulence or anomalies. As such, some businesses are plagued by fraudulence and possible anomalies in their transactions [30]. Telecommunications companies and credit industries are some of the most plagued industry, since it is hard to detect such fraudulence in these businesses. On the other hand, anomalies are not deliberate actions like fraudulence. They are unusual behaviours that may manifest in normal data for unknown reasons, corrupted data, glitches, or network errors in transmissions. If not identified correctly, anomalies can lead to a shift and possible divergence in finding real frauds. Anomalies must be detected and left outside the data set, as to leave out the surge or upheavals these anomalies bring in the data. Fraudulence is however deliberate actions are, usually caused for monetary and financial gains, and is usually carried out by white-collar criminals, insiders, high profile technology criminals, and expert computer hackers.

Usually credit companies try to find certain patterns of charging for its customers. For example, usually stolen credit cards will result in erroneous behaviours and transactions. These transactions usually happen in short periods, within few hours; huge sums of figures in thousands to millions are spent in numerous transactions. The fraud detection used to be manual, through reviewing record manually. As BI fraud detection can easily spot such anomalies. Fraud detection is not only limited to fraudulent customers, but to also fraud within the company, as in fraudulent reports and predictions. BI tools detect suspected behaviours over periods of time, providing what accounts or individuals that require special intention. BI is also used in financial crimes like money laundering, through sifting records of individuals, and insider trading, trading upon secret inside information, Intrusion detection, by outsiders into the system, and spam detection, a major problem facing many enterprises email systems. Finally, we can say that the main problem with such tools would be the false positives that can occur with high percentages in some cases, and miss predictions that have legal and monetary consequences.

Table 1. Summary of BI Advantages

| Business Aspect | Business Intelligence Advantage | Benefits |
|---|---|---|
| Competitive Advantage | • Market Research<br>• Risk Management<br>• Manufacturing Optimization | • Finding Elements of Market Dominance<br>• Bankruptcy Prediction, Better Investments<br>• Better material usage, shipments, scheduling |
| Customer Relationship Management | • Customers' targeting<br>• Pricing Discrimination<br>• Market Baskets<br>• Customers Satisfaction | • Target specific customers with the right products<br>• Dynamic pricing<br>• Better Marketing and Advertisements<br>• Find the reasons and the costs of switching, churn, and satisfactory levels |
| Logistic and Supply Chain Management | • Production Managements<br>• Scheduling Supply Chain<br>• Dynamic Reactions<br>• Forecasting | • Prevent overproduction and underproduction<br>• Help dynamically manage the supplies during their move through the chain<br>• React immediately to changes to help sustain supply<br>• Forecast the demand for production |
| Anomalies and Fraud Detection | • Fraud Detection<br>• Anomaly Detection | • Help find fraudulence transactions, fraudsters, hackers, and possible counterfeiting<br>• Find what data to leave out, why such anomalies happened, and avoid considering them. |

# 6. UTILIZING MINING TO GAIN BUSINESS ADVANTAGES

## 6.1. Jaeger Uses Data Mining to Locate Losses

Jaeger is midsized privately owned British chain cloth retailer in the UK [56] with hundreds of outlets through the different regions of the UK. During the 2008 recession in the UK, shoplifting was on the rise. Retailers like Mark and Spencer, Tesco were accumulating losses in a rate that is at least two times higher than the previous years. According to The Centre for Retail Research, customers stole more than £ 1.6 billion, while employees stole £ 1.3 billion, and finally suppliers took more than £ 209 million fraudulently. The centre has also a Global Retail Theft Barometer, which was estimated at 1.3% of the total loses in sales in the UK in 2007. Current technologies to defend against such theft are the common Closed-Circuit Televisions CCTV surveillance systems, also widely used in retail shops are Electronic Article Surveillance EAS. However, even with such technologies implemented; many retailers still can't estimate losses. The new and innovative technologies made possible that data mining tools, to help in locating the lose source.

Jaeger, in 2008, has implemented a DM application to identify its losses. Their application was centralized, used to process data collected from all the other outlets to try to make sense of it. The DM application for Jaeger called LossManager, provided by IDM software. The software uses feed from Jaeger's other information systems such as the Electronic Point of Sales EPOS system; as it was meant to monitor the employees' behaviour as well as an EPOS. As it is quite common for fraud transactions to be made by some employees in the form of unauthorized discounts. Jaeger did had another DM software before its current solution, but it was far too cumbersome and it did not integrate with the Jaegers' other information systems very well. The current LossManager however was developed with the current systems in mind to provide possible integration.

Jaeger's main aim with LossManager was not to detect shoplifters, as CCTV and the EAS systems were aimed at that purpose. LossManager's aim was to spot theft coming from employees, through fraudulent transactions, loss of inventories, and marginal losses from unnecessary working practices. The audit team at Jaeger was in charge of reading the reports generated from the DM tool. The team analysed the reports to identify the possible dishonest employees and the sources of lost money. However, the report generate by LossManager, led to some false positives. To overcome this, audit team used these reports to investigate further through the data using the DM tool. These questions helped avoid false positives the system generated. Questioning process helped the team to understand how the tool generated reports and what reasons for the patterns did it conceive.

LossManager reduced losses to less than losses accumulated from theft and losses in 2007. Jaeger is already expecting a Return on Investment ROI in its first year of using the system. Although the managers were concerned with the double checking process that the audit team takes in every step. Major findings were that the losses from employees theft consists a very small portion of the losses. In addition, the DM tools found out many erroneous transactions, not all were identified as fraud though. The system also helped Jaeger to better manage its inventory stocks, reducing lose gained from the stock going off-season, or missing items from the other outlets' inventories.

Jaeger experience with LossManager proved that the DM tool, which was meant for fraud and theft detection, did manage to be useful in inventory stock management as well. The DM tool helped Jaeger in 2009, when the recession worsened. The DM tool helped in keeping a low profile on its losses generated from lose and theft. It worth noting that the main difficulty faced was mainly came from the integration with the current systems at Jaeger. The DM tool, LossManager, was implemented in C++, using Microsoft Development Environment, to interface with most other systems. The feeds from the EPOSs, EASs were challenging, as they were the source of info into the DM tool. In addition, the system being centralized made the data collection time consuming as well. As LossManager was integrated well with the EAS feeds, it showed a significant relation between stores with high lose figures and frequent EAS breakdowns.

## 6.2. KFC/Pizza Hut Find a Better BI Tool

KFC/Pizza Hut in Singapore [68], have more than 120 outlets, with a workforce of 5000 employees. As an international fast food franchise, they deliver food and beverages to customers through outlets, drive-through, and by home delivery. To deal with such workload, KFC/Pizza Hut have used a BI tool; the tool was growing increasingly inefficient with each month. Tool didn't meet with time requirements to deliver business reports. It was also had problems with performance benchmarking, plus daily reports across multiple systems was tedious. KFC/Pizza Hut most important daily operation was to calculate payments needed for daily paid workers, such as deliver staff. The used BI tool, managers would take hours and had to work for extra hours to sum up pay correctly. Finally, the old system reporting was slowing down KFC/Pizza Hut ability to match and adapt to current and rapid changes.

Solution was to find another BI tool that was modern. KFC/Pizza Hut contracted with Zap, a BI vendor, using their product Zap Business Intelligence. New solution was web based; it was also linked to other external sources. Corporate data warehouse was remodelled as to include the point of sales POS, marketing, human resources HR, and the corporate very own supply chain. In September 2009, after two month of testing, KFC/Pizza Hut went live with the new BI tool. The employees and managers were generally happy with the new tool. As it was web based, and it offered modern BI capabilities like dashboards, instant report generation, KPI benchmarking, scoreboards, and a very user-friendly interface.

The benefits of the new tool were significant. The improvement included optimized market spending, through live updates; KFC/Pizza Hut immediately responded and adjusted its marketing campaigns and offers. Restaurant planning and outlet location managements were based on reports given for the tool, to cope with KFC/Pizza Hut strategy of being close to its customers. Customer service was highly improved, especially the home delivery service, as the tool accurately capture the parameters of such deliveries to optimize the delivery process. In addition, the POS integration into the data warehouse allowed KFC/Pizza Hut to manage its deals and offers per outlet; different customers at different locations had very varied demands.

Finally, the new BI tool, Zap Business Intelligence, had an expected ROI within 12 months of deployment, as the staff members were reduced, and daily time wasted on reporting was cut to minutes. Workforce efficiency increased and managers do not have to work those extra hours every day. The major cost reduction came from the lessened reliance on the IT staff. The new tool was web based and very intuitive and friendly to use, as most of the operations were carried over to the servers situated at KFC/Pizza Hut's central IT centre. Little IT provision was needed on the different outlets, and fewer staff could manage the new BI tool.

# 7. CHALLENGES FACING MINING TECHNOLOGIES

Many challenges hinder mining tools in business today. They come in three main categories, technological, ethical, and legislative challenges. Mining tools are classified as highly specialized software. This means that they cost in millions, and require extensive infrastructure. The need for this infrastructure stems from their sophisticated and specialized nature, as these tools are considered business professional tools. Aside from this, the human resource, to manage such tools are very scarce, as most business graduates are not trained to use such tools, as most of them are trained on using manual techniques [37].

Technical challenges include huge and elaborate infrastructural needs, and software limitations. Most mining tools need data warehouses as a perquisite with its hardware infrastructure, in case of DM. TM and WM require dedicated hardware and a set of software, the hardware include high-end application server and web servers, distributed computational grids are the main platform for such mining tools. The set of software include, network software tools, NLP packages, even supporting DM packages, and a IR engine to help in search the WWW. These perquisites not only have high costs, but also need expert IT staff as well. Finally, the limitations of such tools today may hinder their usage, as most software package bought from vendors are highly specialized in one single area, like data visualization tools, market analysis tools, for example. In addition, these packages are limited in the sense that they are not extendable; they use certain models, certain techniques that may not be suited for the business models or environment of the procuring enterprise. To a lesser extent, the software interface is limited in some tools or cumbersome, reducing they usage by employees, managers, and executives. Finally, software limitation includes scalability, not all solution scale as well or adaptable to the business environment.

Ethical challenges raised from public concerns about the data found in the Internet. Customers' profiles include private data. Aim of mining tools is not to identify such individuals however, as most data is anonym-zed before use. Still, concerns are raised around how enterprises use individuals' data. Beside the mining application for business, governments are utilizing such mining tools for its national security purposes. Such governmental security agencies try to locate individuals, possible terrorists. These uses, along with business uses of mining tools made public awareness of their legal and privacy rights more evident in programs like Total Information Awareness program [55]. The Total Information Awareness Program was a secret program for the Pentagon, it was aimed at national security and the identification of terrorists, and it used mining tools to sift private individuals' records. Public awareness against the exploitation of individuals' privacy and private data forced the congress to stop funding for this program in 2003. Legal regulations were issued to address these concerns, acts like Health Insurance Portability and Accountability Acts HIPAA, in the United States stated by the congress. The HIPAA act requires a prior consent from individuals regarding the use of their information and the notification of the purpose will their information will be used. Another ethical issue in the mining tools is that they made Globalization far easier. Globalization has dire consequences on emerging economies, as emerging businesses that can never compete with international top companies.

Legislatively BI has resulted in new levels of transparency, due to the vast data decimation across the net willingly or unwillingly, Wikileaks for example. The term data quality [54], is a relatively recent term, refers to authentic, complete, and accurate data and that the source of this data is legally liable for its authenticity. International legislative and professional organizations have made standards and regulations regarding the quality of published data from companies and other agencies. According to [57], more than 25% of critical data in top companies' databases are inaccurate and incomplete. Since BI relies on the data its fed, the quality plays a crucial role, as the quality of the BI is as good as the fed data quality, better and accurate data yields better BI decisions. Open and free market standards today have regulated and insisted that public companies must give accurate fiscal reports, with actual numbers and figures. Such quality of these reports is yet a problem facing BI tools. As it is a problem with taxes collection from such free market enterprises, tax evasion. The transparency of such reports and facts are especially crucial in the energy market. Whereas these energy companies, incredibly and heavily dependent on fiscal reports, cannot only shift market predictions, but also shift other enterprises' plans as well as governments' plans and budget estimates. High quality data is needed for BI in energy markets [52], as demand predictions and market forecasting is the only way to schedule operations and supplies.

## 8. CONCLUSION

Competitiveness today is driven through BI. Companies achieving high competitiveness are the companies utilizing BI tools. Mining technologies had come a long way. The software development, along with hardware developments made possible of more commercially available mining tools. Revolution of information brought in by the Internet and the telecommunications technologies, made them a huge source of information, sometimes for free even.

BI utilizing this vast amounts of data can help in achieve competitive advantages, better customers' relationships, effective resource planning, and fraudulence detection. As BI tools implement AI techniques, decision trees, NLP, and SM technologies, they are considered as sophisticated and highly specialized tools. Many challenges hinder the further developments of such tools. The challenges are technological, ethical, and legislative. As more enterprises and governments are more dependent on such tools, we think that some obstacles have to go in order to progress. As more developments and innovation is coming everyday into the free market, we see a very bright future for BI tools. A long side the information systems that companies are dependent on these days, future companies will depend on BI tools just as much as their information systems.

### ACKNOWLEDGEMENTS

The author would like to thank Professor Hameed Al-Qaheri, College of Business Administration, Kuwait University for his support and help during this research.

**Author : Abdulaziz R. Alazemi**

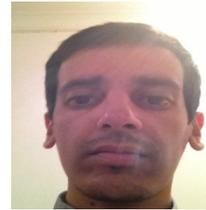

A. R. Alazemi is a graduate researcher interested in the fields of personal information, privacy, networks, and data mining. He graduated from Kuwait University in 2009. He published conference and journal papers regarding PII, data mining and visualization, and distributed systems algorithms.